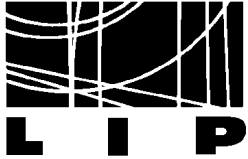 LABORATÓRIO DE INSTRUMENTAÇÃO E FÍSICA EXPERIMENTAL DE PARTÍCULAS

# Perspectives for Positron Emission Tomography with RPCs

A. Blanco[1,2], V. Chepel[1,3], R. Ferreira-Marques[1,3],
P. Fonte[1,4,*], M.I. Lopes[1,3], V. Peskov[5], A. Policarpo[1,3]

1 – LIP – Laboratório de Instrumentação e Física Experimental de Partículas, Portugal
2 – GENP, Dept. Fisica de Particulas, Univ. Santiago de Compostela, Spain.
3 – Departamento de Física da Universidade de Coimbra, Coimbra, Portugal.
4 – Instituto Superior de Engenharia de Coimbra, Coimbra, Portugal.
5 – Royal Institute of Technology, Stockholm, Sweden.

**Abstract**

In this study we address the feasibility and main properties of a positron emission tomograph (PET) based on RPCs. The concept, making use of the converter-plate principle, takes advantage of the intrinsic layered structure of RPCs and its simple and economic construction. The extremely good time and position resolutions of RPCs also allow the TOF-PET imaging technique to be considered.

Monte-Carlo simulations, supported by experimental data, are presented and the main advantages and drawbacks for applications of potential interest are discussed.

---

[*] Corresponding author: Paulo Fonte, LIP-Coimbra, Departamento de Física da Universidade de Coimbra, 3004-516 Coimbra, PORTUGAL.
tel: (+351) 239 833 465, fax: (+351) 239 822 358, email: fonte@lipc.fis.uc.pt

## 1 – Introduction

Positron Emission Tomography (PET) is a radiotracer imaging technique in which tracer compounds labelled with positron emitting radionuclides are injected into the object of study. These tracer compounds can then be used to track biomedical and physiological processes, with applications ranging from the early detection of cancer to neurophysiology studies.

After a short path length the positron annihilates with an electron of the medium emitting simultaneously two almost anti-parallel 511 keV photons. The coincident detection of both photons identifies the occurrence of an annihilation along the chord joining the detection points and the accumulation of such data allows the reconstruction of the activity distribution in the tissues.

If, additionally, the coincidence time difference (difference in flight time between the pair of photons) is measured (TOF-PET) with a FWHM accuracy $\Delta t$ the position of the annihilation along the chord may be identified with a FWHM accuracy $\Delta L$ given by

$$\Delta L = c\Delta t/2 = c(2.36\sqrt{2}\, s_t)/2 \qquad (1)$$

where $s_t$ is the rms time accuracy per photon. The numerical factors combine to yield

$$\Delta L[mm] \approx 0.15 \Delta t[ps] \approx s_t[ps]/2 . \qquad (2)$$

In any foreseeable system the localization accuracy given by eq.(2) is much coarser than the desired image granularity (few mm) and therefore the image reconstruction procedure cannot be avoided. However, by including the TOF information, along with other advantages [1], one gains a sensitivity improvement of the order of $L/\Delta L$, where $L$ is the typical object length, thus reducing the number of events needed for the image reconstruction [2]. For this reason the investigation of TOF-PET systems is being actively pursued (e.g. [3], [4]).

In order to compare a TOF-PET system with a conventional (non-TOF) one and account for the above-mentioned increase in sensitivity when using TOF information, one may introduce an intrinsic sensitivity merit factor defined as

$$f \approx \left(\frac{e_{TOF}}{e}\right)^2 \frac{L[mm]}{0.15\Delta t[ps]} . \qquad (3)$$

where $e_{TOF}$ and $e$ are the quantum efficiencies for a single gamma photon in TOF and non-TOF systems, respectively.

However, in a practical PET tomograph many additional factors determine the overall system sensitivity in terms of dose given to the patient for an exam requiring a certain image quality (see for instance [5], [6] and references therein). Examples of such factors would include:

1) System price, which largely determines the affordable field of view (FOV: axial detector length);
2) Localisation accuracy of the photon interaction points, which includes the depth-of-interaction (DOI) information, the position resolution along the axial and transaxial directions of the tomograph and the errors due to Compton scattering in the detector;
3) The count rate capability, including detector occupancy and data acquisition (DAQ) throughput;
4) The time resolution of the counter and of the DAQ, which influences the number of random coincidences;
5) Counter geometry, in particular the use (2D-PET) or not (3D-PET) of collimating septa dividing the FOV in nearly independent rings.

For most of these performance factors there will be considerable differences between the RPC TOF-PET and the crystal PET approaches. However a full discussion of such aspects and of their interplay cannot be done in the present report, being only outlined the most prominent differences to be expected between these approaches based on simulations and on experimental results.

## 2 – The RPC TOF-PET concept

The RPC TOF-PET concept is based on the converter plate principle [7] and takes advantage of the naturally layered structure of RPCs, of its simple and economic construction, excellent time resolution (60 ps σ even for single gaps equipped with position-sensitive readout [8], [9]) and very good intrinsic position accuracy (50 μm online in digital readout mode [10]).

A possible RPC structure for TOF-PET applications is shown in Figure 1. The incoming gamma photons will interact with the electrode materials through the photoelectric and Compton processes. The resulting electrons strongly scatter in the material and eventually emerge into the gas gap, ionising the gas and initiating avalanches along their path. The exponential dependence of the avalanche final charge with the position of the initial charges assures that only those avalanches initiated

close to the cathode will be detected. (In Figure 1 the lead foil was chosen as cathode.)

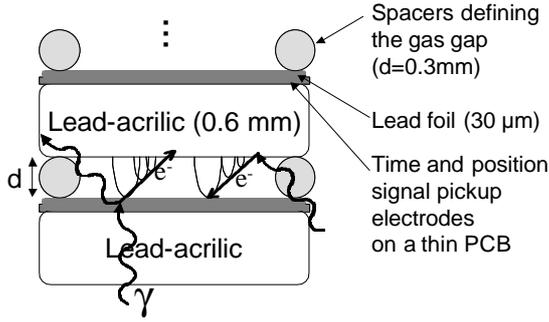

Figure 1 – Counter construction assumed in the GEANT4 simulations presented below with a schematic indication of the most relevant physical processes. Each layer has a total thickness of 1 mm resulting in a 10 cm thick 100-layer counter.

Other approaches based on the converter plate principle, although offering a much poorer time resolution, may be found in [11], [12].

### 3 – Experimental and Monte-Carlo results

*3.1. Quantum Efficiency and Energy Sensitivity*

Clearly the weaker point of the converter plate approach to medium-energy photon detection is the low quantum efficiency (probability of detection per incident photon) relative to inorganic scintillators.

Calculations show that a stack of very thin lead foils obtains the best results. However, for keeping the mechanical accuracy of the counter, the foils must be mounted on a rigid resistive electrode. A possible material may be lead-loaded acrilic (LA - used for transparent X-ray protections), which offers its best performance at a convenient thickness of 0.5 to 1 mm.

We measured a quantum efficiency of 0.0044 for single lead foils in good agreement with ([7], [12]), while the corresponding GEANT4 simulation yielded a value of 0.0042, validating GEANT4 for this application.

Further simulations suggest that a 140-layer counter with the structure shown in Figure 1 should reach an efficiency per photon $\varepsilon_{TOF}$=22% (Figure 2), yielding a relative merit factor eq.(3) for a 24 cm diameter object

$$f \approx \left(\frac{0.2}{0.8}\right)^2 \frac{240}{0.15 \times 300} = 0.33, \qquad (4)$$

where a coincidence time resolution of 300 ps FWHM was used (see section 3.3).

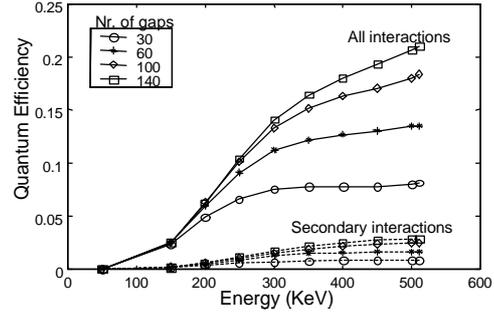

Figure 2 – Simulated (GEANT4) quantum efficiency as a function of the incident photon energy for a stack of counters (see Figure 1). For a large number of plates the values scale almost linearly with the energy above ~100 keV. Only a small fraction of the secondary interactions (from counter-scattered photons) is visible.

The proposed system would show a sensitivity handicap by a factor of three when compared with a standard crystal PET, all other variables being the same. However for some special applications (see section 3.5) there are advantages in the present proposal that may easily outweigh this handicap.

Using photon energy distribution data generated by the public domain SIMSET simulation package [13] it was determined that the energy sensitivity feature of the present approach is equivalent, in terms of rejection of coincidences involving a scattered photon, to an energy discrimination threshold of 300 keV.

*3.2. Position resolution*

A strong point of the converter plate approach is its position resolution capability (e.g. [14]).

In Figure 3 we show the GEANT4-simulated image of a disc phantom in water imaged by a 16-layer counter with a structure similar to that shown in Figure 1. The inner radius was 4 cm and the stack was 16 mm thick. All high-energy processes were simulated including the positron path range (using the *b* energy distribution from $^{18}$F) and photon non-colinearity. The FWHM RPC position resolution was assumed to be twice the width of the gas gap (*2 ´0.3=0.6 mm*) for electrons ejected from the anode and *0.2 mm* for electrons ejected from the cathode. The image was reconstructed by the standard algorithm of filtered backprojection. The contrast of the resulting image was enhanced by subtraction of the low frequency background.

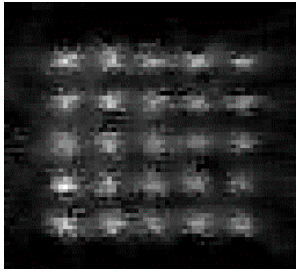

Figure 3 – Monte-Carlo simulation (GEANT4) of 40 μm diameter discs spaced by 1 mm. The discs are clearly resolved, in agreement with the corresponding 0.38 mm FWHM Point Spread Function.

Since the depth of interaction of the photons is milimetrically determined by the depth of the firing gap no parallax effect was observed in the simulations.

The 40 μm diameter discs spaced by 1 mm are clearly resolved. The corresponding Point Spread Function was determined by reconstructing point sources and shows a width of 0.38mm FWHM.

*3.3. Time resolution*

Measurements taken with 0.3 mm single-gap RPCs made with electrodes of glass and aluminium yielded a FWHM coincidence time resolution $\Delta t = 300\,ps$, equivalent to $s_t = 90\,ps$, for 511 keV photon pairs.

Since such counters are known to deliver a resolution $s_t \approx 60\,ps$ [8] when irradiated with minimum ionising particles (MIPs) and no optimisation of the mechanical and operational parameters for this application was done yet, further improvements are reasonably to be expected.

Anyhow, these results already compare favourably with the typical TOF-PET coincidence resolution of 500 ps FWHM achievable with $BaF_2$ or LSO segmented crystals [3] and even with large $BaF_2$ crystals yielding 340 ps FWHM [15].

*3.4. Rate capability*

Since glass (or even bakelite) RPCs are amongst the gas counters with lower rate capability, it should be asked whether these counters might withstand the very large counting rates desirable for PET.

The counting rate capability of timing glass RPCs under continuous irradiation is around *200 Hz/cm²* [16], corresponding to *20 kHz/cm²* for a 100-layer counter. For the large area application considered in the next section, the corresponding maximum singles count rate would be *1.2 GHz*, which is likely unreachable due to other reasons such as random coincidences and DAQ throughput limitations.

Therefore it seems that for large area counters the maximum system counting rate will not be limited by the RPCs rate capability.

*3.5 Field of view and overall system sensitivity*

The present approach may allow the realization of affordable full-body FOV scanners.

Simulations by other authors [17] suggest that increasing the axial FOV of a standard crystal PET from 20 to 60 cm would improve the noise-equivalent count rate (NEC[1] [18]) by a factor of five to eight. Such results are shown in Figure 4 along with our own simulations, based on the SIMSET package [13].

Our curves correspond to a 80 liter water cylinder with a diameter of 24 cm and 1.8 m length. The simulation is meant to determine only geometrical effects and doesn't include rate-dependent effects like system dead time or random coincidences, while the results of ref. [17] where obtained with the Zubal phantom, using larger counting rates and modelling rate-dependent effects. To allow a comparison between both results the curve from [17] was normalized to a NEC sensitivity of 10 kHz/(kBq/ml) for a 20 cm FOV (a common value for 3D-PET: e.g.[19]) and our curve to 3.3 kHz/(kBq/ml) following eq.(4). Two acceptance angles[2] were considered.

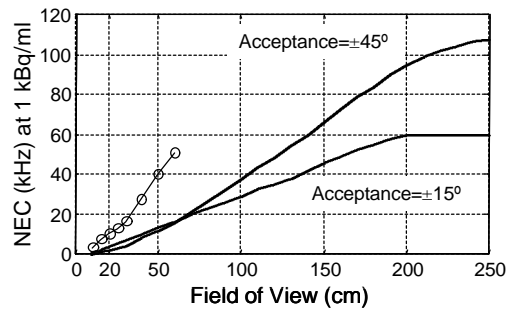

Figure 4 – Simulated noise-equivalent count rates as a function of the axial field of view. The circles correspond to standard 3D PET [17] and the thick lines to RPC TOF-PET. A sensitivity increase by a factor 10 over standard 3D PET (20 cm FOV) may be possible for whole-body examinations.

The results suggest that a 2.5 meter FOV RPC TOF-PET would show a NEC sensitivity improvement by a factor 10 over the standard PET sensitivity, vastly outweighing the intrinsic sensitivity handicap for whole-body examinations.

---

[1] The number of true coincidences that would create an image of similar quality in the absence of noise (scattered and random coincidences).

[2] The maximum angle of the incoming photons relative to the radial direction.

## 4. – Conclusions

We assessed several key features of an economic and positionally accurate RPC-based approach to the TOF-PET medical imaging technique.

Efficiency up to 22% for 511 keV photons may be achievable for a 140-layer counter along with an energy sensitivity equivalent in terms of scatter rejection to an energy cutoff at 300 keV.

A coincidence time resolution of 300 ps FWHM was measured for 511 keV photons, comparing favourably with the values achievable with segmented fast crystals.

The excellent (380 μm FWHM) parallax-free position resolution expected for the reconstructed images (from small diameter counters) may have direct interest for the high-resolution imaging of small animals required by pharmaceutical research.

The simple structure of RPC-type detectors and the consequent possibility of economic construction in large areas opens the way to consider also whole-body field-of-view TOF-PET counters. Simulations suggest that such systems may reach an overall sensitivity one order of magnitude larger than present day crystal-based PET scanners. Such improvements may open new medical applications by lowering considerably the radiation hazard involved in whole-body PET examinations and simultaneously increase the patient throughput.

## 5. – Acknowledgements

This study was supported by the FCT project CERN/P/FIS/40111/2000.

The authors gratefully acknowledge the "Centro de Supercomputación de Galicia" (CESGA) for providing very substantial computational resources and the very special collaboration of Dr. Carmen C. Bueno.

## 6. – References

[1] S. E. Derenzo et al., "Critical instrumentation issues for <2 mm resolution, high sensitivity brain PET," in Quantification of Brain Function: Tracer Kinetics and Image Analysis in Brain PET, K. Uemura, N. A. Lassen, T. Jones, and I. Kanno, Eds. Amsterdam: Elsevier Science Publishers, 1993, pp. 25-37 (http://cfi.lbl.gov/instrumentation/Pubs/QuantBrainFunc.pdf).
[2] T. F. Budinger, J. Nucl. Med., vol. 24, pp. 73-78, 1983.
[3] W. W. Moses and S. E. Derenzo, IEEE Transactions on Nuclear Science NS-46, pp. 474-478 (1999).
[4] T. Yamaya et al., Phys. Med. Biol. 45 (2000) 3125-3134.
[5] W. W. Moses, Nucl. Instrum. and Meth. A471 (2001) 209 –214
[6] D. Crosetto, "A modular VME or IBM PC based data acquisition system for multi-modality PET/CT scanners of different sizes and detector types", presented at the IEEE Nuclear Science Symposium and Medical Imaging Conference, Lyon, France, 2000, published in the conference record. http://www.3d-computing.com/pb/ieee2000-563.pdf.
[7] J.E.Bateman, Nucl. Instr. and Meth 221 (1981) 131.
[8] A. Blanco et al., Nucl. Instr. and Meth. A478 (2002) 170.
[9] A. Blanco et al., "Single-gap timing RPCs with bidimensional position sensitive readout for very accurate TOF systems", these proceedings.
[10] V. Peskov and P. Fonte, "Gain, Rate and Position Resolution Limits of Micropattern Gaseous Detectors", presented at the PSD99-5th International Conference on Position-Sensitive Detectors, London, England, 1999. Also preprint LIP/01-06 (http://xxx. lanl. gov/abs/physics/0106017).
[11] A Jeavons et al., IEEE Trans. Nucl. Sci. NS-30, pp 640-645, 1983.
[12] J.Lacy et al., Nucl. Instr. Meth. A 471 (2001)88.
[13] T.K Lewellen et al., "The SimSET program," in Monte Carlo calculations in nuclear medicine: applications in diagnostic imaging, M. Ljungberg; S-E. Strand, and M. A. King, Eds. Philadelphia: Institute of Physics, 1998, pp 77-92.
[14] A. Jeavons et al., IEEE Trans. Nucl. Sci. Vol 46, No 3, pp 468-473, June 1999.
[15] J.Cederk . et al., Nucl. Instr. and Meth. A471 (2001) 200 -204.
[16] A.Akindinov et al., Nucl. Instr. Meth. A456 (2000) 16.
[17] R.D. Badawi et al., IEEE Trans. Nucl. Sci. NS-47, pp 1228-1232, 2000.
[18] S.C.Strother et al., IEEE Trans. Nucl. Sci. Vol. 37, pp. 783-788, 1990.
[19] Lars-Eric Adam et al, Journal of Nuclear Medicine Vol. 42 No. 12 (2001) 1821-1830